\documentstyle[referee]{mn}

\newcommand{\beq}{\begin{equation}}              
\newcommand{\beqa}{\begin{eqnarray}}              
\newcommand{\eeq}{\end{equation}}             
\newcommand{\eeqa}{\end{eqnarray}}             
\newcommand{\eeqi}[1]{\quad#1\end{equation}} 
\newcommand{\eeqai}[1]{\quad#1\end{eqnarray}} 

\newcommand{\apj}{ApJ}
\newcommand{\mnras}{MNRAS}
\newcommand{\aj}{AJ}

\newif\ifAMStwofonts

\ifoldfss
  \ifCUPmtlplainloaded \else
    \NewTextAlphabet{textbfit} {cmbxti10} {}
    \NewTextAlphabet{textbfss} {cmssbx10} {}
    \NewMathAlphabet{mathbfit} {cmbxti10} {} 
    \NewMathAlphabet{mathbfss} {cmssbx10} {} 
  \fi
  \ifAMStwofonts
    \ifCUPmtlplainloaded \else
      \NewSymbolFont{upmath} {eurm10}
      \NewSymbolFont{AMSa} {msam10}
      \NewMathSymbol{\upi}     {0}{upmath}{19}
      \NewMathSymbol{\umu}     {0}{upmath}{16}
      \NewMathSymbol{\upartial}{0}{upmath}{40}
      \NewMathSymbol{\leqslant}{3}{AMSa}{36}
      \NewMathSymbol{\geqslant}{3}{AMSa}{3E}

      \let\leq=\leqslant \let\le=\leqslant
       
    \fi
  \fi
\fi 

\ifnfssone
  \newmathalphabet{\mathit}
  \addtoversion{normal}{\mathit}{cmr}{m}{it}
  \addtoversion{bold}{\mathit}{cmr}{bx}{it}
  \newmathalphabet{\mathbfit} 
  \addtoversion{normal}{\mathbfit}{cmr}{bx}{it}
  \addtoversion{bold}{\mathbfit}{cmr}{bx}{it}
  \newmathalphabet{\mathbfss} 
  \addtoversion{normal}{\mathbfss}{cmss}{bx}{n}
  \addtoversion{bold}{\mathbfss}{cmss}{bx}{n}
  \ifAMStwofonts
    \ifCUPmtlplainloaded \else
      \UseAMStwoboldmath
      \makeatletter
      \new@mathgroup\upmath@group
      \define@mathgroup\mv@normal\upmath@group{eur}{m}{n}
      \define@mathgroup\mv@bold\upmath@group{eur}{b}{n}
      \edef\UPM{\hexnumber\upmath@group}
      \new@mathgroup\amsa@group
      \define@mathgroup\mv@normal\amsa@group{msa}{m}{n}
      \define@mathgroup\mv@bold\amsa@group{msa}{m}{n}
      \edef\AMSa{\hexnumber\amsa@group}
      \makeatother
      \mathchardef\upi="0\UPM19
      \mathchardef\umu="0\UPM16
      \mathchardef\upartial="0\UPM40
      \mathchardef\leqslant="3\AMSa36
      \mathchardef\geqslant="3\AMSa3E

      \let\leq=\leqslant \let\le=\leqslant

    \fi
  \fi
\fi 

\ifnfsstwo
  \DeclareMathAlphabet{\mathbfit}{OT1}{cmr}{bx}{it}
  \SetMathAlphabet\mathbfit{bold}{OT1}{cmr}{bx}{it}
  \DeclareMathAlphabet{\mathbfss}{OT1}{cmss}{bx}{n}
  \SetMathAlphabet\mathbfss{bold}{OT1}{cmss}{bx}{n}
  \ifAMStwofonts
    \ifCUPmtlplainloaded \else
      \DeclareSymbolFont{UPM}{U}{eur}{m}{n}
      \SetSymbolFont{UPM}{bold}{U}{eur}{b}{n}
      \DeclareSymbolFont{AMSa}{U}{msa}{m}{n}
      \DeclareMathSymbol{\upi}{0}{UPM}{"19}
      \DeclareMathSymbol{\umu}{0}{UPM}{"16}
      \DeclareMathSymbol{\upartial}{0}{UPM}{"40}
      \DeclareMathSymbol{\leqslant}{3}{AMSa}{"36}
      \DeclareMathSymbol{\geqslant}{3}{AMSa}{"3E}

      \let\leq=\leqslant \let\le=\leqslant

    \fi
  \fi
\fi 

\ifCUPmtlplainloaded \else
  \ifAMStwofonts \else 
    \def\upi{\pi}
    \def\umu{\mu}
    \def\upartial{\partial}
  \fi
\fi

\input epsf

\title[]
{Probing the Structure of Lensing Galaxies with Quadruple Lenses:
	The Effect of the ``External'' Shear}
\author[]
	{Hans J. Witt
	\thanks{E-mail: hwitt@aip.de} \\
	Astrophysikalisches Institut Potsdam, An der Sternwarte 16, 
	14482 Potsdam, Germany
	\and
	Shude Mao
	\thanks{E-mail: smao@mpa-garching.mpg.de} \\
	Max-Planck-Institute f\"ur Astrophysik,
	Karl-Schwarzschild-Strasse 1, 85740 Garching, Germany}
\date{Accepted ........
      Received .......;
      in original form .......}
\pagerange{\pageref{firstpage}--\pageref{lastpage}}
\pubyear{1997}

\begin{document}
\maketitle
\label{firstpage}

\begin{abstract}
We study a general elliptical potential of the form
$\psi(x^2+y^2/q^2)~ (0<q\le 1)$ plus an additional shear (with an
arbitrary direction) as models for the observed quadruple lenses.
It is shown that a minimum additional shear is needed even
just to reproduce the observed positions alone.
We also obtain the dependence of the 
axial ratio, $q$, on the orientation of the major axis of potential.
A general relation also exists between the shear, the
position angle and axial ratio of the lensing galaxy. The
relation shows a generic degeneracy in modelling quadruple lenses.
In particular, it shows that 
only the ratio of the ellipticity, $\epsilon\equiv
(1-q^2)/(1+q^2)$,
to the magnitude of shear, $\gamma$ can be determined.
All these results are valid {\it regardless} of the
radial profile of the potential. 
Our formalism applies when the galaxy position is observed,
which is the case for seven of the eight known quadruple lenses.
Application to these seven cases reveals two quadruple lenses
CLASS 1608+656 and HST 12531--2914, requiring highly significant
shear with magnitude $\approx 0.2$. For HST 12531--2914, 
there must be a misalignment between the major axis of light
and the major axis of potential (mass). 
We conclude that detailed modelling of quadruple lenses 
can yield valuable quantitative information about 
the shape of lensing galaxies and their dark matter halos.

\end{abstract}


\section{INTRODUCTION}

Up to now, more than twenty multiply imaged quasars have been
discovered, with roughly half double and half quadruple lenses
(Keeton \& Kochanek 1996, hereafter KK96, see also
Schneider, Ehlers \& Falco 1992,
Blandford \& Narayan 1992 and Kochanek \& Hewitt 1996
for general reviews). These systems, in particular the quadruple
lenses, provide a unique tool
to probe the potentials of galaxies (e.g., Kochanek 1991; Wambsganss
\& Paczy\'nski 1994; Witt, Mao \& Schechter 1995). Recently, 
Keeton, Kochanek \& Seljak (1996) found that while simple
models such as singular isothermal density ellipsoids provide
a reasonable statistical model for the whole sample, no individual
quadruple system is well fitted by such models. They
found numerically that an additional shear term can drastically reduce
$\chi^2$ in fitting while changing the radial distribution
of the potential helps little. As only a limited number of radial
profiles (such as power laws) have been explored numerically,
it is not clear whether the bad fits can still be due to our
incomplete knowledge of the radial profile of galaxies.
In this paper, we study a general class
of elliptical potentials of the form $\psi (x^2 +y^2/q^2)$,
where $q$ denotes the axial ratio of the elliptical potential.
We show that a minimum additional shear is required even if
one is trying to fit only the observed positions. We also
obtain analytical formulae
for the axial ratio and orientations of the potential and shear.
This implies some generic degeneracy in modelling of quadruple lenses.
Our formalism applies {\it regardless} of the functional form of $\psi$,
as long as the lensing galaxy position is observed. 

The results presented here (\S 2) complement the
study by Keeton et al. (1996) and provide
an analytical understanding of their results.
Our analytical formalism makes it possible to check quickly
whether any elliptical potential or density distribution
without shear can work at all without computing $\chi^2$.
We apply the formalism to seven of the eight known
quadruple lenses, including three of the four quadruple lenses 
studied by Keeton et al. (1996).
Our results are consistent with theirs for these three.
For two of the other four cases, HST 12531--2914 and CLASS 1608+656,
we found that the minimum shear required is $\ga 0.2$.
For HST 12531--2914, the major axis of potential must be
misaligned with that of the light. The origin of
the large additional shear and its implications are discussed in 
the last section.

\section{ELLIPTICAL POTENTIAL PLUS SHEAR}

Both the elliptical density and potential distributions are widely used in
the literature to model gravitational lenses 
(e.g., Blandford \& Kochanek 1987, Kochanek \& Blandford 1987,
Kormann, Schneider, \& Bartelmann 1994a,b). These two distributions
resemble each other when the ellipticity is small (Kassiola \& Kovner 1993),
which is generally the case for the known quadruple lenses. We
will use the elliptical potential model due to its analytical
simplicity. Our results should apply to the elliptical
density distribution almost equally well.
To be quadruply lensed by an elliptical potential, the
source must be located closely behind the centre of the lensing
galaxy. For a pure elliptical potential, the possible locations
of the images and the lensing galaxy are highly restricted (Witt 1996,
hereafter W96).
In reality, the image positions and flux ratios will depart
from those predicted by a pure elliptical potential.
For example, large scale structure and/or other galaxies along the line
of sight can distort the image configuration. In addition, any
departure of the galactic potential from the idealized elliptical form
can produce deviations as well. As the pure 
elliptical potential can reproduce the overall observed image configuration
quite well, a reasonable approach for further refinement is to model all the
other perturbations as an additional shear term in the lens equation 
(Kovner 1987). This is the approach we will adopt here,
as in Keeton et al. (1996).

We therefore assume the potential can be modelled
as a two-dimensional elliptical potential plus an additional shear
in an arbitrary direction.
The elliptical potential is by definition given by
$\psi(r_e)$,
where $r_e \equiv x^2 +y^2/q^2$, and
$q$ $(0<q\leq 1)$ is the axial ratio of the potential.
The centre of the galaxy is always located at the origin. 
For clarity, we first assume that the $x$-axis 
coincides with the major axis of the lensing potential.
The results derived in this special coordinate system (which we
term as the major axis frame) are then generalized to
the case with an arbitrary major axis orientation afterwards.
Throughout the paper, all the quantities measured in a general
coordinate system will have a prime superscript
to avoid confusion with those measured in the major axis frame.

The (projected) surface mass distribution is given by
$\Delta \psi = 2 \kappa(x,y)$, where 
$\kappa(x,y)=\Sigma(x,y)/\Sigma_{\rm crit}$
is expressed in units of the critical surface mass density $\Sigma_{\rm crit}$
which depends on the distances to the deflector and the source
(cf. Schneider et al. 1992).
The two-dimensional deflection angle is then simply given by
the derivatives of the potential,
{\boldmath $\alpha$}$=${\boldmath $\nabla$}$\psi$
plus the two terms related to the shear.
The lens equation can be written as
\beqa \xi &=& x + \gamma_1 x + \gamma_2 y
         - \frac{\partial \psi(x^2+y^2/q^2)}{\partial x} 
         = x +\gamma_1 x +\gamma_2 y - 
               \frac{\partial \psi(r_e)}{\partial r_e} 2 x,
 \label{xi} \\  \label{eta}
\eta &=& y +\gamma_2 x -\gamma_1 y
       - \frac{\partial \psi(x^2+y^2/q^2)}{\partial y }
      = y +\gamma_2 x- \gamma_1 y
    -\frac{\partial \psi(r_e)}{\partial r_e} \frac{2 y}{q^2}, 
\eeqa
where $(\xi, \eta)$ denotes the (unknown) source position and
the magnitude of the shear is given by
$\gamma=\sqrt{\gamma_1^2+\gamma_2^2}$.
The shear can also be written in a ``vector'' form
$(\gamma_1, \gamma_2)
=(\gamma \cos2\theta_\gamma, \gamma \sin2\theta_\gamma)$ (
$0\le\theta_\gamma<\pi)$. Notice
that the factor of 2 before $\theta_\gamma$ arises because the shear
is not really a vector but a tensor.
When the shear is acting on-axis, we have $\theta_\gamma=0, \gamma_2=0$.
The shear is maximum off-axis when $\gamma_1=0$, i.e., when
$\theta_\gamma=45^\circ$, or $135^\circ$.

For quadruple lenses, the positions of the four images obey
eqs. (\ref{xi}) and (\ref{eta}), therefore for each of the four images
we can eliminate the factor $\partial \psi(r_e)/\partial r_e$ to 
obtain the following equation:
\beq \label{xiyi} 
\frac{\xi-x_i-\gamma_1 x_i-\gamma_2 y_i}{\eta-y_i-\gamma_2
x_i +\gamma_1 y_i} 
= q^{2} \frac{x_i}{y_i}
\quad \mbox{for} \quad i=1,...,4.
\eeq
In the next three subsections, we will use eq. (\ref{xiyi}) as basis
to derive analytical results in this paper.

\subsection{Lower Limit On the Additional Shear} 

Using the four equations as in eq.(\ref{xiyi}),
we can eliminate first $q$ and then
$\xi$ and $\eta$, which leads us to the following equation:
\beq \gamma_1 a_1+\gamma_2 a_2 +a_3=0,
\quad \label{shear}
\eeq 
where the coefficients $a_1$, $a_2$ and $a_3$ depend on the four 
relative image positions, and are given by
\beqa \label{a1}
a_1 &=& (x_1^2+y_1^2) f_{234} -(x_2^2+y_2^2) f_{341} 
+ (x_3^2+y_3^2) f_{412} - (x_4^2+y_4^2) f_{123}, \\
a_2 &=&- y_1^2 h_{234} +y_2^2 h_{341} - y_3^2 h_{412} +y_4^2 h_{123},
\label{a2} \\
a_3 &=& (x_1^2-y_1^2) f_{234} - (x_2^2-y_2^2) f_{341}
+ (x_3^2-y_3^2) f_{412} - (x_4^2-y_4^2) f_{123},  \label{a3}
\eeqa
with the functions $f_{ijk}$ and $h_{ijk}$ defined as
\beqa \label{fijk} 
f_{ijk} &=& x_i y_i [ x_j y_k - x_k y_j] 
+ x_j y_j [x_k y_i - x_i y_k] + x_k y_k [x_i y_j -x_j y_i]\\
h_{ijk} &=& x_i^2 [ x_j y_k - x_k y_j] \label {hijk}
+ x_j^2 [x_k y_i - x_i y_k] + x_k^2 [x_i y_j -x_j y_i].
\eeqa
Note that $f_{ijk}$, $h_{ijk}$, $a_1, a_2$, and $a_3$ are all odd under 
the permutation of any two indices.

When we derived eq. (\ref{shear}) we assumed that the major axis
of the lensing potential is along the $x$-axis. In practice, it
is usually more difficult to measure the orientation of the galaxy
than to measure its position and those of the images.
In any case, what we measure is the position angle of the light
distribution, not that of the
potential (mass) which enters the lens equation. Hence,
we need to generalize eq. (\ref{shear}) to the case when the major
axis of the galaxy potential is unknown, i.e.,
we need to investigate what happens to the coefficients
$a_1$, $a_2$ and $a_3$ when the major axis of the lensing galaxy 
is rotated by an angle $\theta_G (-\pi/2<\theta_G \le \pi/2)$.
The positions of the images in the new coordinate system, $(x_i^\prime,
y_i^\prime)$, are
related to those measured in the major axis coordinate system, $(x_i,
y_i)$, by
\beqa x_i &=& x_i^\prime \cos \theta_G + y_i^\prime \sin\theta_G, \\
      y_i &=&-x_i^\prime \sin \theta_G + y_i^\prime \cos\theta_G,
\eeqa
for $i=1,..,4$.

In the appendix we show that $a_3$ is rotationally invariant, i.e.,
\beq
a_3 = a_3^\prime.\label{a3_transform}
\eeq
Since $\theta_{\gamma}=\theta_{\gamma}^\prime-\theta_G$, 
the shear tensor transforms like
\beqa \gamma_1 &= \gamma \cos(2\theta_\gamma)
=\gamma_1^\prime \cos (2\theta_G) + \gamma_2^\prime \sin(2\theta_G), \\
      \gamma_2 &= \gamma \sin(2\theta_\gamma)
=- \gamma_1^\prime \sin (2\theta_G) + \gamma_2^\prime \cos(2\theta_G).
\eeqa
To satisfy the invariance of $a_3$, $(a_1, a_2)$ must transform like
a tensor as well. It is easy to verify that $(a_1, a_2)$ 
transforms as follows
\beqa
a_1 &=& a_1^\prime\cos(2 \theta_G) + a_2^\prime \sin(2\theta_G), 
\label{a1_transform}\\
a_2 &=& -a_1^\prime \sin(2 \theta_G) +a_2^\prime \cos(2\theta_G).
\label{a2_transform}
\eeqa
Substituting eqs. (\ref{a3_transform}), (\ref{a1_transform}), and
(\ref{a2_transform}) into eq. (\ref{shear}),
one obtains
\beq
\gamma_1^\prime a_1^\prime +  \gamma_2^\prime a_2^\prime + a_3^\prime = 0.
\eeq
The equation has the same form as eq. (\ref{shear}), but now
all the quantities are evaluated in the general coordinate system.
Replacing the two shear components by
$\gamma_1^\prime =\gamma \cos(2\theta_{\gamma}^\prime)$ and
$\gamma_2 =\gamma
\sin(2\theta_{\gamma}^\prime)$ yields
\beq \gamma[ a_1^\prime \cos(2\theta_\gamma^\prime)+a_2^\prime \sin(
2\theta_\gamma^\prime) ] +a_3^\prime =0.
\eeq
The above equation can be rewritten as
\beq \gamma \sqrt{{a_1^\prime}^2+{a_2^\prime}^2} \sin(2\theta_\gamma^\prime
+\varphi^\prime) +a_3^\prime
=0,
\label{rotate}
\eeq
where $\varphi^\prime~ (-\pi < \varphi^\prime \le \pi)$ 
is the polar angle of the vector $(a_2^\prime, a_1^\prime)$. 
Eq. (\ref{rotate}) implies that, to fit the 
observed positions, a minimum shear is required:
\beq
\gamma_{\min} =
\frac{\vert a_3^\prime \vert}{\sqrt{{a_1^\prime}^2 +{a_2^\prime}^2}}=
\frac{\vert a_3 \vert}{\sqrt{{a_1}^2 +{a_2}^2}},
\quad \mbox{when} \quad \theta_{\gamma}^\prime =
\theta_{\gamma,{\rm min}}^\prime \equiv -{\varphi^\prime \over 2}
- {\pi \over 4} \mbox{sign}(a_3^\prime) + k \pi,
\label{minimum}
\eeq
where throughout the paper $k$ is
an integer that make the angle at the left hand side of the equation
($\theta_\gamma^\prime$ here) fall into the right range.

We now make some general remarks about eq. (\ref{minimum}).
First, the minimum shear is required no matter
what the radial profile is as long as the iso-potential contours are
ellipses. This easily explains why changing radial profiles, such as
adding a core radius or changing the slope of a power law radial
profile, will not improve the fitting much (Kochanek 1991;
Wambsganss \& Paczy\'nski 1994; Keeton et al. 1996).
Second, as the coefficients
$a_1, a_2$ and $a_3$ involve differences of permuting terms, the 
required (minimum) shear is likely to be very sensitive to the accuracy of
positions. High quality relative astrometry of galaxy and image 
positions are thus much desirable.

\subsection{Axial Ratio of Lensing Galaxy}

In this subsection we discuss whether the
axial ratio $q$ can be restricted. To do this,
we again start with eq.(\ref{xiyi}) and eliminate successively
$\xi$, $\eta$ using three image coordinates $i=1,2,3$, after which
we obtain
\beq \label{qg1g2}
(1+\gamma_1 -q^2(1-\gamma_1)) f_{123} = \gamma_2
[q^2 h_{123} -y_1^2 (x_2 y_3 -x_3 y_2) -y_2^2 (x_3 y_1- x_1 y_3) - y_3^2
(x_1 y_2 -x_2 y_1)].
\eeq
where $f_{132}$ and $h_{132}$ are defined as in eqs. (\ref{fijk})
and (\ref{hijk}). 
If $\gamma_2 \ne 0$, then
we can use the fourth image position to eliminate $\gamma_1$ or
$\gamma_2$. By eliminating one shear component the other
shear component factorizes out of the equation simultaneously. 
Therefore we obtain an equation which depends
only on the relative image positions and the axial ratio:
\beq \label{q2}
q^2= \frac{y_1^2 f_{234} -y_2^2 f_{341} + y_3^2 f_{412} - y_4^2f_{123}}{
x_1^2 f_{234} -x_2^2 f_{341} + x_3^2 f_{412} - x_4^2f_{123}}
={a_1-a_3 \over a_1 + a_3},
\eeq
where $a_1$ and $a_3$ are given by eqs. (\ref{a1}) and (\ref{a3}).
We can easily generalize eq. (\ref{q2}) to an arbitrary major
axis orientation by using eqs. (\ref{a1_transform})
and (\ref{a3_transform}):
\beq \label{q2_general}
q^2={ a_1^\prime\cos(2 \theta_G) + a_2^\prime \sin(2\theta_G) - a_3^\prime \over
a_1^\prime\cos(2 \theta_G) +a_2^\prime \sin(2\theta_G) + a_3^\prime}
= {\sin(\varphi^\prime+2\theta_G)-\mbox{sign}(a_3^\prime)\gamma_{\rm min} \over
\sin(\varphi^\prime+2\theta_G)+\mbox{sign}(a_3^\prime)\gamma_{\rm min}},
\eeq
with $\varphi^\prime$ as defined below eq. (\ref{rotate}).
Since
$q^2$ must be positive, eq. (\ref{q2_general})
provides a general (but weak) constraint on the axis orientation
(cf. Fig. 3).
Clearly $q$ achieves a maximum,
\beq \label{qmax}
q_{\rm max} = \left(1-\gamma_{\rm min}\over 1+\gamma_{\rm min}\right)^{1/2},
\quad \mbox{when} \quad
\theta_G = \theta_{G,{\rm max}} \equiv -{\varphi^\prime \over 2} +
{\pi \over 4} \mbox{sign}(a_3^\prime)
+ k \pi .
\eeq

As shown in W96, it is impossible
to determine $q$ for a pure elliptical potential with
a shear acting along the axis, i.e., when $\gamma_2=0$;
it therefore seems rather peculiar 
that for the more general shear case it is actually possible to do so.
The reason for this peculiarity is rooted in the properties of an pure 
elliptical potential. Due to its highly symmetric and self-similar
shape of the iso-potential contours, the positions of the images
and galaxy are highly restricted. For example, the location of the
lensing galaxy is restricted to a hyperbola-like curve and the image
positions in such a potential must fulfil the identities
$f_{123}=f_{134}=f_{234}=f_{124}=0$ (W96). This can be seen
from eq. (\ref{qg1g2}) by setting $\gamma_2=0$. Then
we must have $[1+\gamma_1 -q^2(1-\gamma_1)] f_{132}=0$, which requires
either $q^2=(1+\gamma_1)/(1-\gamma_1), \gamma_1<0$ or $f_{132}=0$.
If the former condition is satisfied, using eqs. (1) and (2), it is
easy to show that all the images must lie on a straight line. This
image configuration clearly does not
resemble the observed quadruple lenses. Therefore
we must have $f_{123}=0$. As a result $q$ can no longer be determined.
Physically we can understand it as follows: A off-axis shear
on top of the elliptical potential breaks
down the high symmetry required for the image positions ($f_{ijk}$'s
are no longer required to be zero), which in turn allows
us to determine the axial ratio. 

As we have shown, when the shear is acting on-axis, eq. (\ref{q2_general})
cannot be applied. Numerically this implies that when the shear component
($\gamma_2$) is small,
eq. (\ref{q2_general}) is likely to be unstable due to the
errors in the image and galaxy positions.
Therefore before applying eq. (\ref{q2_general}) it is necessary to check
first
whether the observed system requires an off-axis shear,
using the test introduced by W96 (see \S 2.4).
If a significant
off-axis shear is indicated, the (minimum) magnitude of 
the additional shear can be estimated using eq. (\ref{minimum}).
If the required shear is large,
then eq. (\ref{q2_general}) can be applied to restrict the axial
ratio and the orientation of the major axis of potential (see below).

\subsection{Orientations of Lensing Galaxy and Shear}

Combining eqs. (\ref{rotate}) and (\ref{q2_general}), we arrive at
a new relation:
\beq \label{new}
\tan(2\theta_G + \varphi^\prime) = - {\sin 2\theta_\gamma
\over \cos 2\theta_\gamma + \epsilon/\gamma}\, , ~~
\epsilon \equiv { 1-q^2 \over 1+q^2} \, ,
\eeq
where $\epsilon$ as defined is the ellipticity of the
iso-potential contours. This equation bears some similarity to
eq. (22) found by Keeton et al. (1996).

Eq. (\ref{new}) shows two important points. First,
since only the ratio $\epsilon$ and $\gamma$ enters eq. (\ref{new}),
there is a degeneracy between these two parameters.
This can be understood as follows:
an increase in ellipticity $(0\leq \epsilon < 1)$ 
(a decrease of 
axial ratio of $q$) stretches the images more along the
$x$ axis, while an increase in the shear stretches the
images more along the $y$ axis, the balance between
these two competing effects introduces the degeneracy.
Eq. (\ref{new}) also shows that even if the
ellipticity and the galaxy orientation are known (e.g., if we
use those for the light distribution), the shear tensor still
cannot be determined uniquely (see 
Figs. 8 and 9 in Keeton et al. 1996).

If the potential is purely elliptical,
i.e., $\theta_\gamma=0$ $(\gamma_2=0)$, then from
eq. (\ref{new}), we have
\beq \label{pure_new}
\theta_{G, {\rm pure}}=-{\varphi^\prime \over 2} \quad \mbox{for}
\quad \gamma_2=0.
\eeq
The above expression can be shown to be identical to
eq. (7) in W96 in this case. However, if an off-axis shear
is present $(0<\gamma_2 \la 0.3)$, eq. (\ref{pure_new})
is only approximately true. Combined with eq. (\ref{qmax}),
we obtain
\beq \label{diff}
\vert \theta_{G,{\rm max}} - \theta_{G,{\rm pure}} \vert
\approx {\pi \over 4}.
\eeq
Notice that in general, the 
difference between the
true orientation and that obtained by using a pure elliptical
potential depend on not only the direction of shear
but also the ratio of ellipticity and shear as well.

\subsection{Application To Known Quadruple Lenses}

Eight quadruple lenses are known (see KK96
for a thorough summary). Seven of these eight systems
(except H1413+117) have known galaxy and image positions;
our results can applied to study these systems. When multiple sets of
positions are available, we generally took the data with the best
astrometry. Detailed references are listed in the last column in
Table 1. For each system, we use eq. (\ref{minimum}) to
derive the minimum shear; the results are shown in Table 1. As one can see,
the required minimum shears vary significantly
from system to system. For example,
for 2237+0305 it is consistent with zero, while
for the lens CLASS 1608+656 it is as large as 0.25.
How sensitive are these estimates to the positional errors?
To address this question, we used Monte Carlo simulations 
to generate synthetic lensed systems by assuming
all the positions are uncorrelated and their errors are Gaussian. 
For each lens, 10,000 Monte Carlo realizations are generated, and
the minimum shear is calculated for each of these. The average
and standard deviation are then computed and shown in Table 1.
In some cases, the distributions of the inferred minimum shear
are highly non-Gaussian, especially in the
case of B1422+231. It has an almost flat distribution of minimum
shear between 0 to $\approx 0.3$. For MG 0414+0534 the minimum shear
is about 0.1 ($2\sigma$ significant). However, it remains unclear whether
the shear is due to the less accurate galaxy 
position \footnote{The Ellithorpe (1995) image position (see KK96)
would imply a considerable minimum shear ($\approx 0.16$) for MG 0414+0534. 
However, his relative position
of image C is not compatible with the recent observations
of Falco et al. (1996) and Katz et al. (1996).}.
Two cases, CLASS 1608+656 (Meyers et al. 1995, hereafter M95)
and HST 12531--2914 (Ratnatunga et al. 1995, 1996, hereafter R95, R96)
require even more significant shears than 
MG 0414+0534. We will therefore concentrate on these two systems below.


\begin{table}
\caption{Minimum Shear Required For Known Quadruple Lenses}
\begin{tabular}{lllll}
    \multicolumn{5}{c}{}\\
    \hline \hline
    &&&&\\
  Object & $\gamma_{\min}$ & $\sigma_{\rm image}$ & 
$\sigma_{\rm galaxy}$ & Reference \\
    &&&&\\
    \hline
    &&&&\\
2237+0305 & $0.0092~(0.0099\pm 0.0063)$ & 0.005 & 0.005 & C91     \\
PG 1115+080 & $0.053~(0.058\pm 0.034)$ & 0.005 & 0.05 & K93  \\
MG 0414+0534 & $0.12~(0.12\pm 0.058)$  & 0.0003 & 0.05 &
KMH96; F96 \\
CLASS 1608+656 & $0.25~(0.25 \pm 0.027)$ & 0.01 & 0.01 & M95; S95  \\
B 1422+231 & $0.11~(0.165 \pm 0.125)$ & 0.002 & 0.05 & P92; YE94
 \\
HST 14176+5226 & $0.037~(0.040 \pm 0.027)$ & 0.03 & 0.03 & R95; R96   \\
HST 12531--2914 & $0.18~(0.18 \pm 0.058)$ & 0.03 & 0.03 & R95; R96 \\
&&& \\
     \hline
&&&\\
  \end{tabular}

\medskip

Note.--- $\gamma_{\rm min}$ is the minimum shear which is 
required to fit the positions of the observed image and lensing galaxy. The
mean and standard deviation (listed in brackets)
for each system in $\gamma_{\rm min}$
are derived from 10,000
Monte Carlo simulations for the positions assuming the positional errors
are Gaussian and uncorrelated.
$\sigma_{\rm image}$ and $\sigma_{\rm galaxy}$ denotes the
astrometric accuracy (in arcseconds) of the image positions and the
lensing galaxy (cf. KK96).
One known quadruple lens H1413+117 is not listed
here because no galaxy position is yet available, therefore our formalism
does not apply.
\end{table}

The image configurations of these two systems, 
HST 12531--2914 and CLASS 1608+656 are shown in Figs. 1 and 2,
respectively. For a pure elliptical potential, the lensing galaxy
position must lie on the hyperbolic-like curves in these figures.
Both systems show a significant deviation these curves,
suggesting the presence of a significant off-axis shear (cf. W96).
This is consistent with the highly significant minimum shear derived above
(cf. Table 1).

\begin{figure}
\epsfxsize=14cm \epsfbox{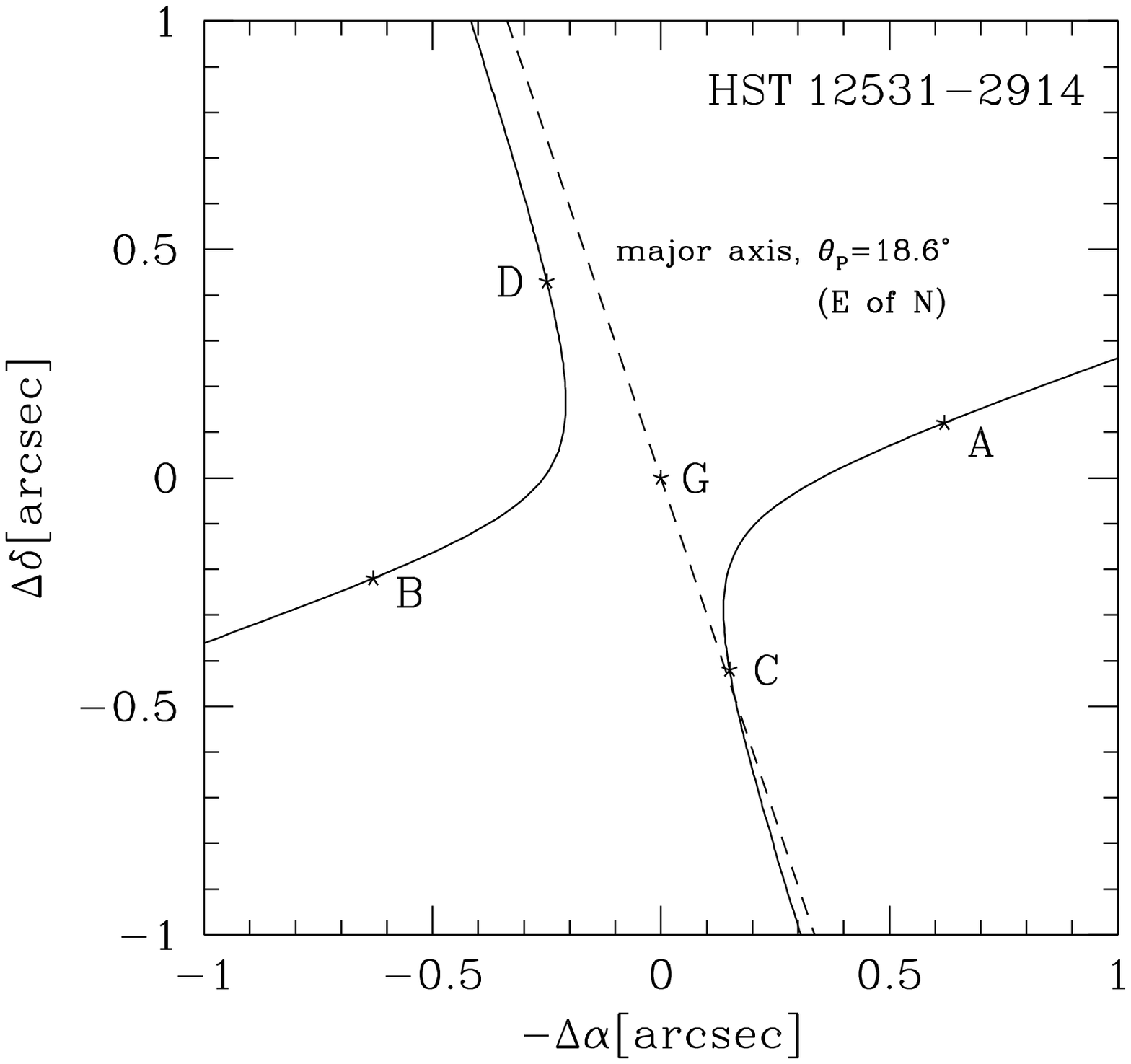}
\caption{Image configuration of HST 12531--2914.
North is up and east is to the left. The errorbar in the galaxy 
position is about 0.03 arcsecond (cf. Table 1). The solid line indicates
th possible location of the lensing galaxy when we assume
a pure elliptical potential. 
For this model, the galaxy and image positions lie on the 
same curves. The offset of the galaxy position from the predicted curves
indicates the presence of a large shear. The expected position angle,
$\theta_P$,
measured from north through east,
for a pure elliptical potential is indicated with a dashed
line (cf. W96).
The observed position angle for the light distribution
is $22^\circ.6\pm 0.5$ from the HST image in the F606W filter (R95; R96).
}
\end{figure}

\begin{figure}
\epsfxsize=14cm \epsfbox{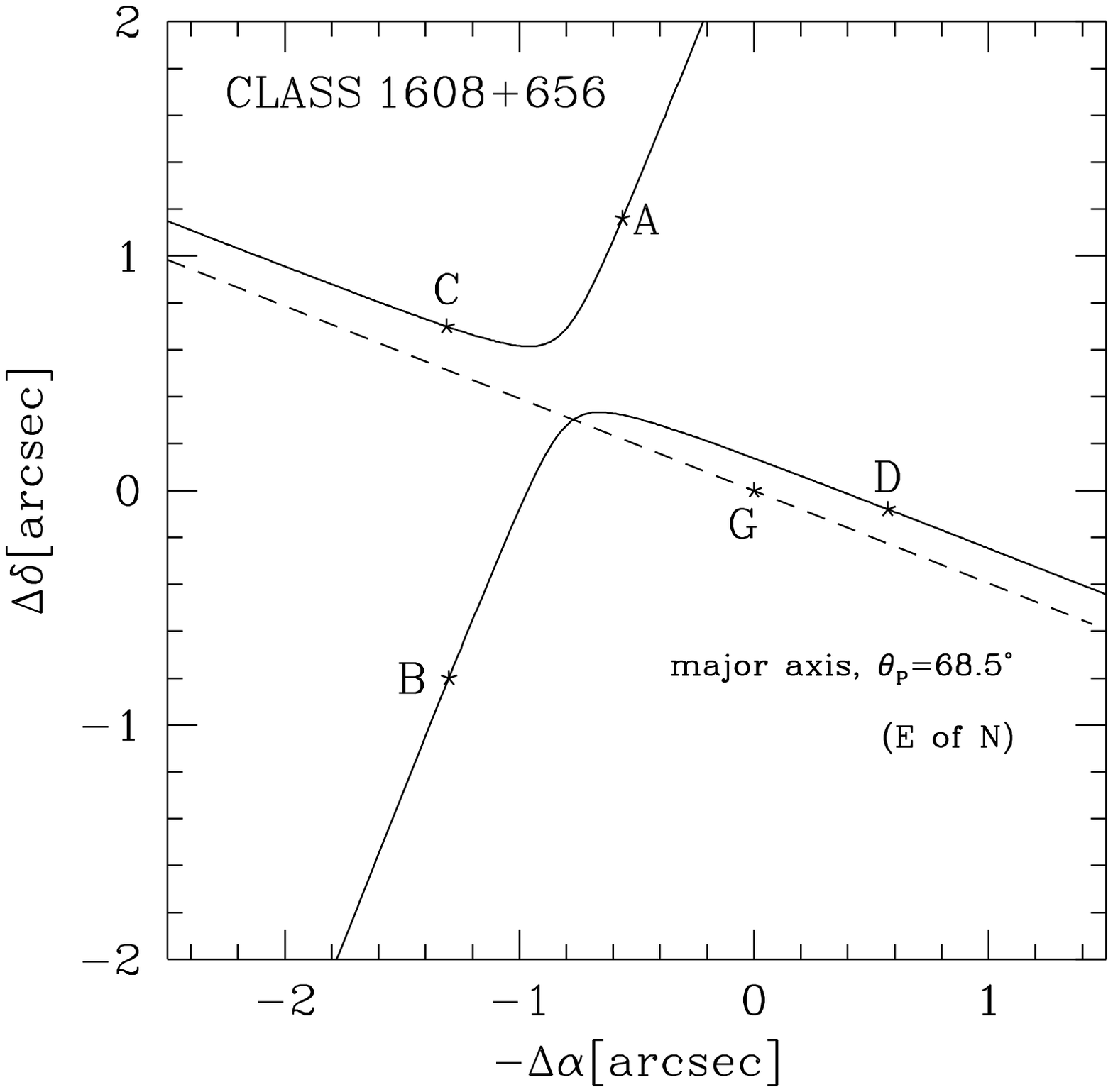}
\caption{Image configuration of CLASS 1608+656.
North is up and east is to the left. The errorbar in the
galaxy position is about 0.01 arcsecond (cf. Table 1). The solid line indicates
the possible location of the galaxy when we assume a pure elliptical 
potential. For this model, the galaxy and image positions lie on the 
same curves. The offset of the galaxy position from the predicted curves
indicates the presence of a large shear. The expected position angle,
$\theta_P$, for a pure elliptical potential is
indicated as a dashed line (cf. W96).
The inferred position angle from modelling is $67^\circ$ (E of N) (M95).
}
\end{figure}

As HST 12531--2914 and CLASS 1608+656 clearly
need some large off-axis additional shear,
eq. (\ref{q2_general}) can therefore be applied to both systems.
In Fig. 3, we plot the dependence of the
axial ratio $q$ on the position angle $\theta_P$ for
these two systems. Note the position
angle is conventionally measured from north (positive $y$ axis) through
east (minus $x$ axis). If we limit the range of $\theta_P$ to be
from 0 to $\pi$, then it is related to $\theta_G$ by
\beq \label{GP}
\theta_P  = \theta_G+\pi/2.
\eeq
If we (artificially) fit a pure elliptical potential model
to both systems, we obtain position angles
$\theta_P=18.6^\circ, \theta_P=68.5^\circ$ for
HST 12531--2914 and CLASS 1608+656, respectively (cf. Fig. 3).
For HST 12531-2914, this estimate is in rough agreement
with the position angle of the light,
$\theta_P=22^\circ.6\pm 0.5$ (R95). (For CLASS 1608+656, the
orientation of the galaxy is unavailable.)
There are a few interesting
things that can be seen from Fig. 3. 
First for HST 12531--2914,
the positional angle of light (solid line)
falls in the unphysical region where $q^2$ is negative.
This means that there must be
a misalignment between the major axis of the potential
and that of the light. This conclusion is valid regardless of
the radial profile of the potential. From Fig. 3, if the
axial ratio of the potential is identical to that of light
(0.73), then $\theta_P \approx 39.5^\circ$, implying a misalignment
of about $17^\circ$.
Second, the $q$ vs. $\theta_G$ curves are
fairly flat around the peak regions. This can be seen by by expanding
eq. (\ref{qmax}) in Taylor
series around $\theta_{G, {\rm max}}$, which leads to
\beq \label{qexpand}
q^2 = q_{\rm max}^2  - {4 \gamma_{\rm min} \over (1+\gamma_{\rm min})^2}
\delta \theta_G^2, ~~ \theta_G = \theta_{G,\rm{max}} + \delta \theta_G.
\eeq
This means that $q^2$ varies quadratically at the peak region.
Third, although we showed that HST 12531--2914 and CLASS 1608+656 cannot
be exactly fitted by a pure elliptical potential, nevertheless
the models published so far did use the pure elliptical
density distributions (R95, R96; M95). In these modelling,
one finds the best fit parameters by minimizing a $\chi^2$ measure.
Obviously the resulting $\chi^2$ per degree of freedom
will be very large (see $\chi^2$ in Keeton et al. 1996 for other systems).
As one typically searches for the best fit
axial ratio starting from an initial guess of one, as shown in Fig. 3
the $q$ vs. $\theta_G$ curve
is nearly flat around the maximum (implying a large phase space
in the multiple dimensional parameter space), we therefore expect 
the resulting axial ratio to be close to the maximum axial ratio .
Indeed, the
axial ratios for the density distribution, $q_\rho$, are found to be
0.37 for HST 12531--2914 (R95; R96) and 0.28 for
CLASS 1608+656 (M95). Since $q \approx 2/3+q_\rho/3$ 
(Binney \& Treimaine 1987),
we have $q=0.83$ and $0.79$. 
These are very close to the
maximum axial ratios from Fig. 3, $q_{\rm max}=0.83$
for HST 12531--2914 and $q_{\rm max}=0.78$ for
CLASS 1608+656, achieved 
at $\theta_P=65.3^\circ$ and $114.2^\circ$, respectively.
Fourth, the position angle at the maximum axial ratio
is approximately
$45^\circ$ away from the orientation inferred by modelling
the lens galaxy as a pure elliptical potential for both systems,
just as given by eq. (\ref{diff}).

\begin{figure}
\epsfxsize=14cm \epsfbox{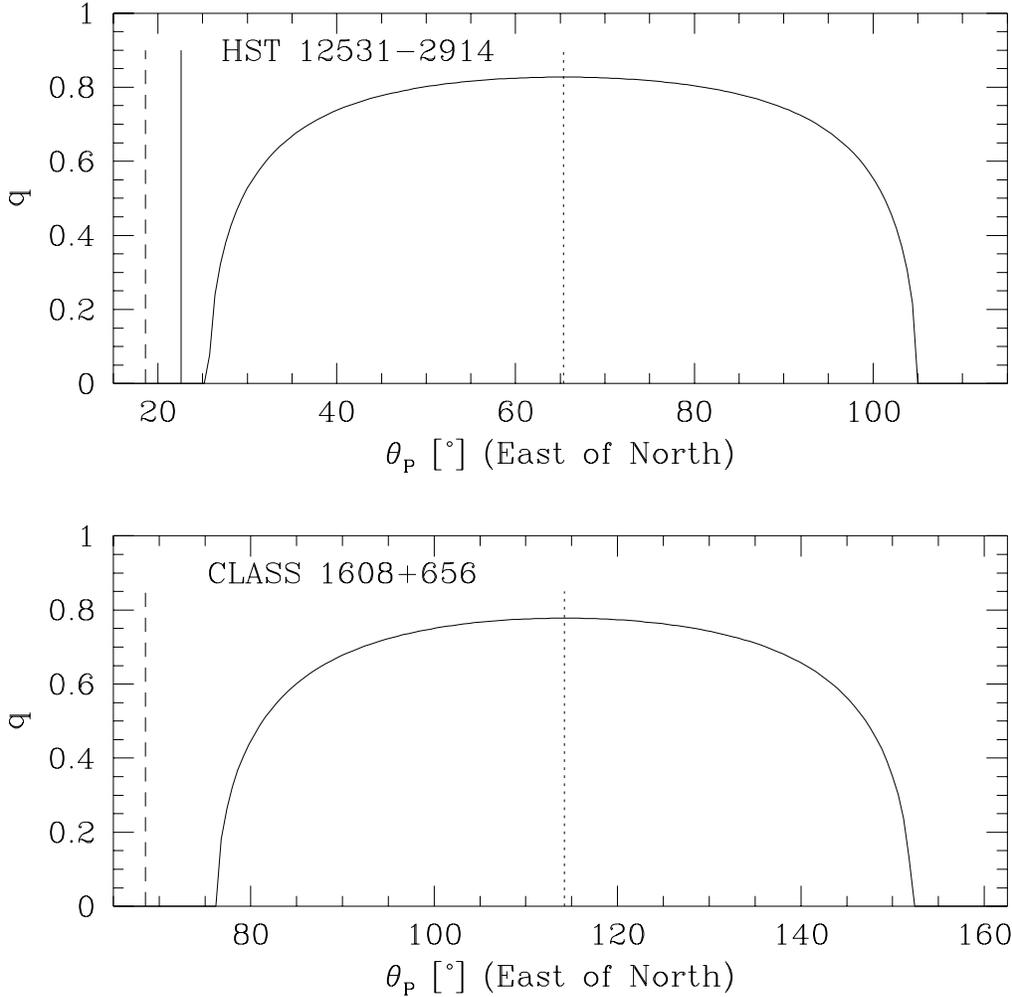}
\caption{Dependence of the axial ratio $q$ on the position angle $\theta_P$
for CLASS 1608+656 (bottom panels) and HST 12531--2914 (top panels). 
The position angle of the light distribution for 
HST 12531--2914 in the F606W filter is shown
with the solid line. The dashed line shows the 
position angle of the major axis
estimated by enforcing a pure elliptical potential while
the dotted line shows the position angle for the maximum $q$
(see eq. [\ref{qmax}]). $q_{\rm max}=0.78$ for CLASS 1608+656
and $q_{\rm max}=0.83$ for HST 12531--2914.
}
\end{figure}

\section{DISCUSSION}

We have studied a general class of models with an elliptical potential plus
an additional shear. It was shown that to fit the image and galaxy positions
in quadruple lenses, the additional shear has to exceed some minimum value.
In addition, an analytical expression for the axial ratio is derived.
We also showed that there is a complex relation between the
orientations of the shear and potential, the ellipticity and
the magnitude of shear. As only the ratio of the ellipticity and
the magnitude of shear enters the relation, these two parameters
are linearly degenerate.
We emphasize our results are valid no matter
what the radial profile is for the elliptical potentials. Applying the
analytical results to seven of the eight known quadruple lenses,
we found that two (MG 0414+0534, B 1422+231) are 
consistent with the presence of additional shears of the order of 0.1,
while HST 12531--2914 and CLASS 1608+656 require highly
significant shears of $\ga 0.2$. 
For HST 12531--2914, we conclude that the major axes of potential
and light must be misaligned, regardless of the detailed potential shape.
We caution that both systems are somewhat ``special'':
HST 12531--2914 has very small separations between the images.
The images seem to be not perfectly aligned with axes of
the lensing galaxy (cf. the frames in R96).
In contrast the
lensed source in CLASS 1608+656 is a radio galaxy and not a quasar.
It is conceivable that the small separations or the extended source
size make the position measurements more difficult and the errors
on their positions are under-estimated. We therefore artificially
enlarged their errors by a factor of 2, and recomputed their
statistical significance. CLASS 1608+656 remains $5\sigma$
significant while the significance for HST 12531--2914 has dropped
to $1.7\sigma$. The exceptional nature of CLASS 1608+656
can also be seen from the ratio of the distances between the images C and
D to A and B, $\Delta\theta_{CD}/\Delta\theta_{AB}\approx 1.0$ (for
HST 12531--2914 $\Delta\theta_{CD}/\Delta\theta_{AB} \approx$0.73).
For a pure elliptical isothermal 
sphere, we would expect 
$\Delta\theta_{CD}/\Delta\theta_{AB} \approx q$.
CLASS 1608+656 therefore signals a substantial
deviation even from the maximum possible value ($q=0.78$), again indicating
the presence of an additional shear on top of an elliptical potential.

We emphasize here that in Table 1 only the lower bound of the additional
shear was derived. Computations with some test potentials
indicates that $\gamma_{\rm min} \approx \gamma_2$ and $q_{\rm max} 
\approx q$ when $\gamma_1 \approx 0$, i.e., when
$\theta_\gamma \approx 45^\circ$, or $135^\circ$. In these
cases the additional shears derived are highly significant.
However, if $\gamma_1 \ga \gamma_2$
the minimum shear derived using eq. (\ref{minimum})
is usually smaller than the actual
shear $\gamma$ and not statistically significant. 
Since for CLASS 1608+656 and HST 12531--2914, the additional shears
are highly significant,
their directions of shear are likely to be close to
$45^\circ$ and the shear is close to the minimum values derived here.
For B 1422+231 and PG 1115+080, Keeton et al. (1996)
gives the additional shears of 0.20 and 0.09, roughly a factor of
2 larger than the predicted minimum shears, suggesting there
is a significant part of shear acting along the axes. The real
shears in other systems
can also be easily a factor 2 or 3 larger than the predicted values.

These additional shears required are large
compared with the external shears produced by
large scale structure (Bar-Kana 1996, Schneider 1997), or
galaxies and clusters along the line of sight (Keeton et al. 1996),
which are usually of the order of a few percent. This
suggests that the origin of the shear is not ``external'', but 
introduced by the lensing galaxy internally (Keeton et al. 1996). If
the shear is truly intrinsic, it will be interesting to see whether
there is correlation of
the additional shear with physical parameters of the lenses. For example,
the potential of galaxies will be less relaxed and more irregular
in a hierarchical formation scenario as the redshift increases.
Therefore when a sufficient number of lenses are
known, one should find some correlation between the required 
additional shear and the redshift of the lensing galaxy.

It is worthwhile to reflect why we need such large additional
shears to model some of these observed systems.
For the elliptical potential studied here (and similarly for the 
elliptical density distribution), the direction of
the deflection angle is independent of the radial coordinates (cf.
eqs. [1] and [2]).
This strongly restricts the allowed image and
galaxy positions (W96). Violation of these restrictions 
directly translates to an additional off-axis shear required in the model.
The large values of the inferred additional shears 
illustrates that our modelling of the lens potential
is too simplistic. The misalignment of the
mass and luminous part of the lensing galaxy in HST 12531--2914
shows an example of possible complexities. Other possibilities
such as the triaxiality of the dark halo clearly exist (see Keeton et al.
1996). All these complications
make the isopotential contours more complex and possibly twisted.
With the added complexities, one can presumably fit the
observed positions and flux ratios better as more degrees
of freedom become available. 
Such complexities may prove to be a nuisance in lens applications
such as determining the Hubble constant. On the other
hand, this means detailed modelling of quadruple lenses
may also yield information about the shapes of
lensing potentials and dark halos. For example,
the newly discovered quintuple lens 0024+1654 
(cf. Colley, Tyson \& Turner 1996) would be an excellent example to apply
our formalism as the fifth image provides additional constraints.
This exceptional case shows a lensed high-redshift galaxy
with large image separations. Since the source is also extended, the system
contains much more information that can be used to
probe the potential of the lensing cluster. With more
and more quadruple (or quintuple) systems discovered, we are
optimistic that gravitational lenses will become an increasingly 
discriminating tool to study the (dark) matter distribution in galaxies.

This work was supported by a postdoctoral grant of the Deutsche
Forschungsgemeinschaft (DFG) under Gz. Mu 1020/3-2 (HJW) and by
the ``Sonderforschungsbereich 375-95 f\"ur Astro-Teilchenphysik'' der
Deutschen Forschungsgemeinschaft (SM). 
We are very grateful to Peter Schneider for his constructive
comments on the paper.

\appendix

\section{Rotational Invariance of $a_3$}

In this appendix, we show that $a_3$ as defined in eq. (\ref{a3}) is
rotationally invariant. It is easy to verify that
$a_3$ can be written as the determinant of a 4x4 matrix:
\beq
a_3 = \det \left(  \begin{array} {cccc}
	x_1^2-y_1^2 & x_2^2-y_2^2 & x_3^2-y_3^2 & x_4^2-y_4^2 \\
	x_1 y_1 & x_2 y_2 & x_3 y_3 & x_4 y_4 \\
	x_1 & x_2 & x_3 & x_4 \\
	y_1 & y_2 & y_3 & y_4
	\end{array}
\right).
\eeq
 From linear algebra, this determinant can be expressed as
\beq
a_3 = \sum_{{i,j,k,l}}
	\epsilon(i,j,k,l) ~
	\det 
	\left(
	\begin{array}{cc}
	x_i^2-y_i^2 & x_j^2-y_j^2 \\
	x_i y_i & x_j y_j
	\end{array}
	\right)
	\times
	\det
	\left(
	\begin{array}{cc}
	x_k & x_l \\
	y_k & y_l 
	\end{array}
	\right),
\eeq
where $(i,j,k,l)$'s are permutations of the four indices (1,2,3,4)
satisfying $i<j$ and $k<l$, and $\epsilon(i,j,k,l)$'s are either
$+1$ or $-1$ but are of no importance here, since we will show each
term in the sum is rotationally invariant.
To prove this,
let us express the image positions in the
polar coordinates, i.e., $(x_i, y_i)=(r_i\cos\theta_i, r_i\sin\theta_i)$.
The determinants of the two 2x2 submatrices are then
\beq
	\det 
	\left(
	\begin{array}{cc}
	x_i^2-y_i^2 & x_j^2-y_j^2 \\
	x_i y_i & x_j y_j
	\end{array}
	\right) = {1 \over 2} r_i^2 r_j^2 \sin 2(\theta_j-\theta_i),~
	\det
	\left(
	\begin{array}{cc}
	x_k & x_l \\
	y_k & y_l 
	\end{array}
	\right) = r_k r_l \sin(\theta_l-\theta_k).
\eeq
Both terms are obviously invariant under rotation. It follows that
$a_3$ is rotationally invariant as well.


{}

\bsp
\label{lastpage}

\begin{thebibliography}{}

\bibitem{B96} Bar-Kana, R. 1996, ApJ, 468, 17
\bibitem{BK92} Blandford, R.D. \& Kochanek, C. S. 1987, ApJ, 321, 658
\bibitem{BN92} Blandford, R.D. \& Narayan, R. 1992, ARA\&A, 30, 311 
\bibitem{BT87} Binney, J., \& Treimaine, S. 1987, Galactic Dynamics
               (Princeton: Princeton University Press), p60
\bibitem{Cr91} Crane, P. et al. 1991, ApJ, 369, L59 (C91)
\bibitem{Co96} Colley, W.N., Tyson, J.A. \& Turner, E.L. 1996, 
                 ApJ, 461, L83 
\bibitem{El95} Ellithorpe, J.D., 1995, Ph.D. thesis, MIT
\bibitem{FLS96} Falco, E.E., Leh\'ar, J., Shapiro, I.I., \&
           Kristian, J. 1996, ApJ, submitted (F96)
\bibitem{Ka96} Katz, C. A., Moore, C. B., \& Hewitt, J. N. 1996,
	preprint (astro-ph/9609104) (KMH96)
\bibitem{KK93} Kassiola, A. \& Kovner, I. 1993, \apj, 417, 450
\bibitem{KKS96} Keeton II, C.R., Kochanek, C. S., \& Seljak, U. 1996,
	preprint (astro-ph/9610163)
\bibitem{KK96} Keeton II, C.R. \& Kochanek, C.S. 1996, 
           in Kochanek, C.S. \& Hewitt, J.N. (eds.),
           IAU 173, Melbourne, {\it Astrophys. Applications 
           of Gravitational Lensing}, Kluwer, p. 419 (KK96)
\bibitem{Ko91} Kochanek, C.S. 1991, \apj, 373, 354
\bibitem{KB92} Kochanek, C. S. \& Blandford, R. D. 1987, ApJ, 321, 676
\bibitem{KH96} Kochanek, C.S. \& Hewitt, J.N. (eds.),
           IAU 173, Melbourne, {\it Astrophys. Applications 
           of Gravitational Lensing}, Kluwer, p. 419
\bibitem{KSB94a}  Kormann, R., Schneider, P., \& Bartelmann, M.:  1994a,
                  A\&A, 284, 285
\bibitem{KSB94b}  Kormann, R., Schneider, P., \& Bartelmann, M.:  1994b,
                  A\&A, 286, 357
\bibitem{K87} Kovner, I. 1987, \apj, 316, 52
\bibitem{Kr93} Kristian, J. et al. 1993, \aj, 106, 1330 (K93)
\bibitem{My95} Myers, S.T. et al. 1995, \apj, 447, L5 (M95)
\bibitem{Pa92} Patnaik, A.R. et al. 1992, \mnras, 259, 1$_{\rm P}$ (P92)
\bibitem{Ra95} Ratnatunga, K.U., et al. 1995, \apj, 453, L5 (R95)
\bibitem{Ra96} Ratnatunga, K.U., et al. 1996, 
           in Kochanek, C.S. \& Hewitt, J.N. (eds.),
           IAU 173, Melbourne, {\it Astrophys. Applications 
           of Gravitational Lensing}, Kluwer, p. 323 (R96)
\bibitem{Sc95} Schechter, P.L. 1995, private communication in KK96 (S95)
\bibitem{S97} Schneider, P. 1997, in preparation
\bibitem{SEF92} Schneider, P., Ehlers, J., \& Falco, E. E. 1992, 
                  {\it Gravitational Lenses} (Springer-Verlag: New York)
\bibitem{WP94} Wambsganss, J. \& Paczy\'nski, B. 1994, \aj, 108, 1156
\bibitem{Wi96} Witt, H.J. 1996, \apj, 472, L1 (W96)
\bibitem{WMS95} Witt, H.J., Mao, S., \& Schechter, P.L. 1995, \apj, 443, 18
\bibitem{YE94} Yee, H.K.C., \& Ellingson, E. 1994, \aj, 107, 28 (YE94)

\end{thebibliography}
\end{document}